\documentclass[%
 reprint,
showpacs,preprintnumbers,
showkeys,
 amsmath,amssymb,
 aps,
prl
]{revtex4-1}
\usepackage{graphicx}
\usepackage{dcolumn}
\usepackage{bm}
\usepackage{geometry}
\usepackage{wrapfig}
\usepackage{mathtools}
\usepackage{amsmath}
\usepackage{amsfonts}
\usepackage{mathrsfs}
\usepackage{subcaption}
\usepackage{hyperref}
\usepackage{multirow}
\usepackage{color}
\usepackage{tikz-feynman}
\tikzfeynmanset{compat=1.1.0}
\usepackage{feynmp}
\usepackage{tikzsymbols}
\usepackage{xfrac}
\usepackage{array}
\usepackage{textcomp} 
\usepackage{phaistos} 

\usepackage{natbib}
\setcitestyle{sort&compress,numbers}
\begin{document}
\title{Radiative neutrino mass via fermion kinetic mixing}
\author{Sin Kyu Kang}
\email{skkang@seoultech.ac.kr}
 \altaffiliation{School of Liberal Arts, Seoul-Tech, Seoul 139-743, Korea}
  \author{Oleg Popov}
  \email{opopo001@ucr.edu}
\altaffiliation{Institute of Convergence Fundamental Studies, \\ Seoul National University of Science and Technology, \\Seoul 139-743, Korea}
\date{\today}

\begin{abstract}
We propose that the radiative generation of the neutrino mass can be achieved by incorporating the kinetic mixing of fermion fields which arises radiatively at one-loop level. As a demonstrative example of the application of the mechanism, we present the particular case of the Standard Model extension by $U(1)_D$ symmetry. As a result, we show how neutrino masses can be generated via a kinetic mixing portal instead of a mass matrix with residual symmetries responsible for the stability of multicomponent dark matter.
\end{abstract}
\maketitle

In the Standard Model (SM) of electroweak (EW) interactions neutrinos are predicted to be massless. There are many extensions of SM that generate Majorana neutrino masses via the Weinberg dimension-five operator~\cite{Weinberg:1979sa} $LHLH/\Lambda$ realized at tree level~\cite{Mohapatra:1979ia,Minkowski:1977sc,Yanagida:1979ab,GellMann:1979grs,Schechter:1980gr,Magg:1980ut,Cheng:1980qt,Mohapatra:1980yp,Foot:1988aq}, and one-loop level~\cite{Zee:1980ai,Ma:2006km,Pilaftsis:1991ug,Grimus:1989pu,Fraser:2014yha,Ma:1998dn}, as well as Dirac neutrino masses~\cite{Ma:2016mwh,Farzan:2012sa}.

In this paper, we introduce a new possibility that kinetic mixing of fermions occurs radiatively and it leads to the radiative generation of
neutrino masses. In order to present the idea of kinetic mixing of fermions that  induces the radiative generation of neutrino masses, we introduce the minimal set of fields in the context of $U(1)_D$ dark  gauge symmetry extension of the SM. The field content for the $SU(3)_c\otimes SU(2)_L\otimes U(1)_Y\otimes U(1)_D$ gauge symmetry case is shown in Table~\ref{tab:U1cont1}, where all fermions are presented by left-handed fields. The last column in Table~\ref{tab:U1cont1} shows the number of copies beside the flavor count. 
The quantum number assignments for these fields are presented in Table~\ref{tab:U1cont1}.
\begin{table}
\caption{Particle content for the $\mathcal{G}_{SM}\otimes U(1)_D$ model.}
\label{tab:U1cont1}
\centering
\begin{tabular}{ccccccc}
\hline
\hline
Field & SU(3)$_c$ & SU(2)$_L$ & $U(1)_Y$ & $U(1)_D$ & \tiny{Flavors} & \tiny{copies} \\ \hline
$Q$ & \textbf{3} & \textbf{2} & $\frac{1}{6}$ & 0 & 3 & 1 \\
$u^c$ & $\Bar{\textbf{3}}$ & \textbf{1} & $-\frac{2}{3}$ & 0 & 3 & 1 \\
$d^c$ & $\Bar{\textbf{3}}$ & \textbf{1} & $\frac{1}{3}$ & 0 & 3 & 1 \\
$L$ & \textbf{1} & \textbf{2} & $-\frac{1}{2}$ & 0 & 3 & 1 \\
$e^c$ & \textbf{1} & \textbf{1} & 1 & 0 & 3 & 1 \\
$H$ & \textbf{1} & \textbf{2} & $\frac{1}{2}$ & 0 & 1 & 1 \\
$A_L$ & \textbf{1} & \textbf{1} & 0 & 3 & 3 & 1 \\
$C_L$ & \textbf{1} & \textbf{1} & 0 & 1 & 3 & 5 \\
$N_L$ & \textbf{1} & \textbf{1} & 0 & $-4$ & 3 & 1 \\
$N_R^c$ & \textbf{1} & \textbf{1} & 0 & 4 & 3 & 1 \\
$S_{L}$ & \textbf{1} & \textbf{1} & 0 & $-2$ & 3 & 4 \\
$\Psi_L$ & \textbf{1} & \textbf{1} & 0 & $\frac{5}{2}$ & 3 & 1 \\
$\Psi_R^c$ & \textbf{1} & \textbf{1} & 0 & $-\frac{5}{2}$ & 3 & 1 \\
$\eta_L$ & \textbf{1} & \textbf{2} & $-\frac{1}{2}$ & 3 & 1 & 1 \\
$\eta_D$ & \textbf{1} & \textbf{1} & 0 & $-1$ & 1 & 1 \\
$\phi$ & \textbf{1} & \textbf{1} & 0 & 2 & 1 & 1 \\
$s_{7}$ & \textbf{1} & \textbf{1} & 0 & $\frac{7}{2}$ & 1 & 1 \\
$s_{11}$ & \textbf{1} & \textbf{1} & 0 & $-\frac{11}{2}$ & 1 & 1 \\ \hline \hline
\end{tabular}
\end{table}

\begin{figure}[!h]
\centering
\includegraphics[scale=1,trim={5cm 22cm 10cm 4cm},clip]{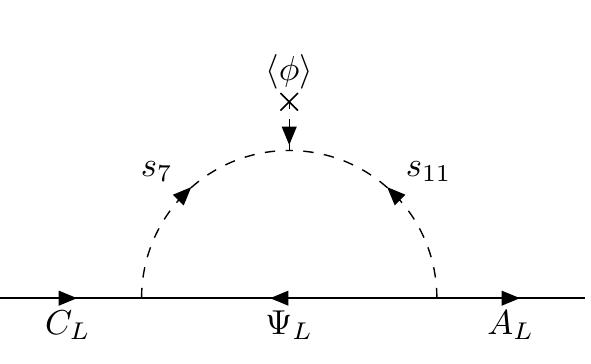}
\caption{Radiative kinetic mixing between $C_L$ and $A_L$ in the $G_{SM}\times U(1)_D$ gauge symmetry case.}
\label{fig:U1KM1}
\end{figure}

First of all, let us show how the kinetic mixing between two new fermions $A_L$ and $C_L^i$ 
can be generated. In order for the kinetic mixing to occur radiatively, we need  mediators which are a massive  fermion $\Psi_L$ and new scalars  $s_{7,11}$. In addition, a scalar  $\phi$ is introduced so as to spontaneously break $U(1)_D$ and then $<\phi>=v_{\phi}/\sqrt{2}$.
The new interaction terms leading to the kinetic mixing between $A_L$ and $C_L^i$ are given by
\small{
\begin{align}
-\mathcal{L}_{KM}&=\Psi_{La}Y^{ab}_{A}A_{Lb}s_{11}+\Psi_{La} Y^{ab}_{C^i}C_{Lb}^is_{7}^*\nonumber \\
&+\mu_3 \phi s_{11} s_7 +\text{h.c.}
\label{eq:lagA}
\end{align}
}
Note that  all terms which could give $A_L$ a mass, even at one-loop order, are forbidden by symmetry, whereas the mass terms for $C_L^i$ are allowed. The Feynman diagram representing the fermion kinetic mixing mechanism is shown in Fig.\ref{fig:U1KM1}. In general, all five copies of $C_L$ can  kinetically mix with $A_L$, but without loss of generality and for the sake of simplicity, we take the bases of $C_L$ fermions in which only one particular $C_L^i$ mixes with $A_L$. 
It is worthwhile to notice that $\phi$ carries even $Q_D$ charge in order to generate a residual $\mathbb{Z}_2$ symmetry upon breaking of $U(1)_D$, which stabilizes the dark matter (DM) candidate. This residual symmetry is desirable to prevent the collapsing of the loop down to tree level, i.e., to prevent the $s_{7}$ and $s_{11}$ from obtaining the vacuum expectation values(VEVs). In our case, it is achieved by choosing the specific $Q_D$ charge assignments for the fields in the loop. The other new fields are required to generate neutrino masses and for cancellation of chiral anomalies, as will be shown later. The result of Fig.~\ref{fig:U1KM1} produces the effective kinetic mixing between two fermion fields which  leads to the Lagrangian kinetic term $\imath a \Bar{A}_L \not{\partial}C_L^i+\text{H.c.}$, where $a$ represents the loop structure and will be given in Eq.~(\ref{eq:a1}).
All relevant kinetic terms are presented by
\begin{align}
\mathscr{L}&=\imath \left(\begin{matrix}
\Bar{A}_L & \Bar{C}_L^i
\end{matrix}\right)\not{\partial}\left(\begin{matrix}
1 & a \\
a^* & 1
\end{matrix}\right)\left(\begin{matrix}
A_L \\ C_L^i
\end{matrix}\right).
\end{align}
In order to bring the kinetic terms into canonical form, the first step is to rotate by $\pi/4$ so that the kinetic matrix becomes diagonalized. Next,  renormalization of the corresponding fermion fields is required for the kinetic terms to be properly normalized. Finally, we need to diagonalize back the mass matrix of the fermions  $A_L$ and $C_L^i$  in the new basis of properly normalized mass eigenstates $F_{1L}$ and $F_{2L}$. This rotation will differ from $\pi/4$ due to the  presence of rescaling. The relation between ($A_L$, $C_L^i$) and ($F_{1L}$, $F_{2L}$) is given by
\begin{align}
\left(\begin{matrix}
A_L \\ C_L^i
\end{matrix}\right)&=U(\pi/4,\delta)^{\dagger}R^{-1}U(\alpha,0)^{\dagger}\left(\begin{matrix}
F_{1L} \\ F_{2L}
\end{matrix}\right)\\
&=\left(\begin{matrix}
1 & -\frac{\epsilon}{\sqrt{1-\epsilon^2}} \\
0 & \frac{\epsilon}{\sqrt{1-\epsilon^2}}
\end{matrix}\right)\left(\begin{matrix}
F_{1L} \\ F_{2L}
\end{matrix}\right).
\end{align}
Here, $U(\theta,\delta)$ is the unitary 2$\times$2 transformation with mixing angle $\theta$ and relative phase change $\delta$, $\text{Im}[\epsilon]=\text{Im}[ae^{-\imath 2\delta}]=0$,  $\text{tan}2\alpha=-\sqrt{1-\left|\epsilon\right|^2}/\left|\epsilon\right|$, and $R=\text{Diag}(\sqrt{1-\left|\epsilon\right|},\sqrt{1+\left|\epsilon\right|})$. $a$ is defined as
\small{
\begin{align}
\label{eq:a1}
a&=\frac{1}{16 \pi^2}Y_A^* \left[s_{sR}c_{sR}G\left(y_{s2R},y_{s1R}\right)\right.\\
&\left.+s_{sI}c_{sI}G\left(y_{s1I},y_{s2I}\right)\right] Y_C, \text{ and}\nonumber \\
&G\left(y_i,y_j\right)=\frac{\left[y_i(y_i/2-1)\text{ln}y_i-y_j(y_j/2-1)\text{ln}y_j\right.}{\left(1-y_i\right)^2\left(1-y_j\right)^2}\nonumber\\
&+y_i y_j(y_i y_j+4)\text{ln}\left(\frac{y_i}{y_j}\right)/2\nonumber \\
&\left.+\frac{1}{2}\left\{y_i(y_i-1)-y_j(y_j-1)+y_i y_j(y_j-y_i)\right\}\right],
\end{align}
}
where $y_i=m_{s_i}^2/M_{\Psi}^2$, and Yukawa couplings are given with the flavor indices suppressed. Here, $M_{\Psi}$ is the Dirac mass of $\Psi_L$, and $s_{sR(sI)}$ and $c_{sR(sI)}$ stand for the sinus  and cosinus, respectively,  corresponding to the mixing between the real (imaginary) components of $s_7$ and $s_{11}$, which is proportional to $\mu_3$ in Eq.(\ref{eq:lagA}). The final form of the relevant Lagrangian is $\imath \Bar{F}_{1L} \not{\partial}F_{1L}+\imath \Bar{F}_{2L} \not{\partial}F_{2L}-M_{F_2}F_{2L}F_{2L}$, and the mass eigenvalues are $M_{F_1}=0$ and $M_{F_2}=m_C \text{ exp}[\imath 2\delta]/(1-\left|\epsilon\right|^2)$, where $m_C$ is the mass of $C_L^i$.

\begin{center}
\begin{figure}[!h]
\centering
\includegraphics[scale=1,trim={5cm 21.5cm 10cm 4cm},clip]{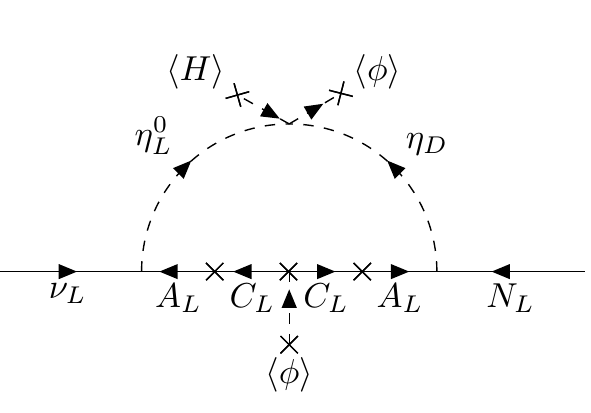}
\caption{Radiative neutrino mass generation via fermion kinetic mixing in the $G_{SM}\otimes U(1)_D$ gauge symmetry case. Crosses between the $A_L$ and $C_L$ fields correspond to kinetic mixing given in Fig.~\ref{fig:U1KM1}.}
\label{fig:U1NUM1}
\end{figure}
\end{center}

Now, let us consider how Dirac neutrino mass can be radiatively generated. For our aim,  lepton doublet $L$ couples only to $A_L$, and  $N_L$ is needed as a Dirac mass partner for $\nu_L$. A fermion field $\Psi_R$ is added to produce a Dirac mass for $\Psi_L$. Also, new scalar fields $\eta_L$ and $\eta_D$ are added for the generation of Dirac \emph{radiative} neutrino mass (see Fig.~\ref{fig:U1NUM1}). The new interaction terms leading to the radiative generation of Dirac neutrino mass are given by
\small{
\begin{align}
\label{eq:lag1}
-\mathcal{L}_{DM}&=L_{ai} Y_{L}^{ab} A_{Lb} \eta^{\dagger i}_L+A_{La} Y^{ab}_{N}N_{Lb}\eta_{D}^* \\
&+C_{La}^i  Y^{ab}_{C^i C^j}C_{Lb}^j \phi^*+\mu_{D} \eta_D^2 \phi \nonumber \\
&+\lambda_{H\eta\phi}H_i \eta_{Lj} \eta_D \phi^* \epsilon^{ij}+\lambda_{s\eta} \eta_D^2 s_{7}^* s_{11}^*+\text{h.c.},  \nonumber
\end{align}
}
where $H$ develops a nonzero VEV,  $\left\langle H^0\right\rangle=v/\sqrt{2}$. Note that  the $\lambda_{H\eta\phi}$ quartic scalar term which mixes $\eta_L^0$ with $\eta_D$, is needed to generate Dirac radiative neutrino mass. Figure~\ref{fig:U1NUM1} represents the radiative neutrino mass generation via kinetic mixing for $G_{SM}\times U(1)_D$ gauge symmetry. Thus, neutrino masses are generated via the effective dimension-eight LH$N_L \phi^2 (\phi^*\phi)$ operator at three-loop order after H and $\phi$ develop nonzero vacuum expectation values after spontaneous symmetry breaking.

The neutrino Dirac mass corresponding to the mixing between {\bf $\nu_L$ }and $N_L$ [see Eq.~(\ref{eq:mnu1})]
 is given by
\begin{align}
\label{eq:mloop}
m_{loop}&=\frac{Y_N M_{F_2}Y_L \epsilon^2}{16\pi^2}\left[s_{\eta R} c_{\eta R} F(x_{1R},x_{2R})\right.\nonumber \\
&\left.+s_{\eta I} c_{\eta I} F(x_{2I},x_{1I})\right],
\end{align}
where $s_{\eta R(\eta I)}$, $c_{\eta R(\eta I)}$ stand for the sinus and cosinus, respectively, corresponding to the mixing between the real (imaginary) components of $\eta_L^0$ and $\eta_D$. $F(x_i,x_j)$ is defined as
\begin{align}
\label{eq:f1}
F(x_i,x_j)&=\frac{x_i \text{ln}x_i-x_j \text{ln}x_j+x_i x_j \text{ln}\frac{x_j}{x_i}}{\left(1-x_i\right)\left(1-x_j\right)},
\end{align}
with {\bf $x_{i}=\frac{m_{\eta_i}^2}{M_{F_2}^2}$}. The extra suppression from $\epsilon$ in $m_{loop}$ allows for a wider range of masses, mass splittings, and Yukawa couplings.

For completeness of the model with $\mathcal{G}_{SM}\otimes U(1)_D$ gauge symmetry, we require that $SU(3)_C\times SU(2)_L\times U(1)_Y$ triangular anomalies are canceled in the same way as in the canonical SM case. Since there are no fermions that transform nontrivially under the SM and dark sector simultaneously, any cross anomalies between the SM and $U(1)_D$ are trivially absent. The only anomalies to consider for cancellation are $U(1)_{D\text{Grav}}$ and $[U(1)_D]^3$. For this purpose, multiple copies of $C_L$ as well as new fermions $S_{L}^i (i=1,..,4)$, $N_R^c$, and $\Psi_R^c$ are introduced. Considering $U(1)_D$ sector anomalies, they cancel in the following way:$\sum_i Q_{Di}=1\times (3) + 5\times(1)+4\times(-2)+1\times(-4)+1\times(4)+1\times\left(\frac{5}{2}\right)+1\times\left(-\frac{5}{2}\right)=0$, $\sum_i Q^3_{Di}=1\times (3)^3 + 5\times(1)^3+4\times(-2)^3+1\times(-4)^3+1\times(4)^3+1\times\left(\frac{5}{2}\right)^3+1\times\left(-\frac{5}{2}\right)^3=0$. In addition to  $\mathcal{L}_{SM}$, $\mathcal{L}_{KM}$, and $\mathcal{L}_{DM}$, we introduce invariant mass terms of  $\Psi_{L(R)}$ and $N_{L(R)}$ and extra new interaction terms  given by
\small{
\begin{align}
-\mathcal{L}_{extra}&=A_{La}Y^{ab}_{AS^i}S_{Lb}^i\eta_D+C_{La}^iY^{ab}_{C^iS^j}S_{Lb}^j \eta_D^* \nonumber \\
&+\bar{S}_{La}^i Y^{ab}_{NS^i}N_{Rb}\phi+\text{h.c.}
\label{eq:lag2}
\end{align}
}
where $C_L^i$ and $S_{L}^j$  run over 1$-$5 and 1$-$4, respectively.
 
After EW and dark symmetry breaking due to the symmetry and field content of the model, we obtain two residual dark $\mathbb{Z}_2$ symmetries which are \emph{not ad hoc}. The first $\mathbb{Z}_2$ symmetry is analogous to the one from the canonical scotogenic model, but here is it obtained from $U(1)_D$ spontaneous symmetry breaking. The other $\mathbb{Z}_2$ symmetry is new and present here due to fractional charge assignments of the particles involved in the kinetic mixing. This gives us the opportunity for the multicomponent DM case. SM fields, $N_{L,R}$, $S_{L}$, and $\phi$ fields transform trivially, fields with $(-1)^{Q_D(=\text{odd})}$ which are $\eta_L, \eta_D, A_L, C_L^i$ transform as $(-,+)$, scalar fields with fractional $Q_D$ charges, i.e., $s_{7}, s_{11}$, transform as $(+,-)$, and $\Psi_{L,R}$ transform as $(-,-)$ under both $\mathbb{Z}_2$ symmetries, respectively.

Let us present the mass spectrum of the new fields we introduced.\\
Fermions:
We need to consider three different sectors that do not mix with each another. First,  the $\left(\nu_L, N_L, N_R^c, S_{L}\right)$ sector, which is $\left(+,+\right)$ under  $\mathbb{Z}_2^{1,2}$. Next, the $\mathbb{Z}_2^1\sim\left(-\right)$ odd sector similar to the one present in the canonical scotogenic paper~\cite{Ma:2006km}, $\left(A_L, C_{L}^i \right)$ fields. Lastly, the $\mathbb{Z}_2^2\sim\left(-\right)$ odd fermions, special for this model, due to the presence of kinetic mixing, i.e., {\bf $\Psi_{L(R)}$}. Starting with the $\mathbb{Z}_2$ even fermions we have and considering the Lagrangian given in Eqs.~(\ref{eq:lag1}) and (\ref{eq:lag2}), we get the following mass matrix for these new fermions and neutrinos
\begin{equation}
\left(\begin{matrix}
0 & m_{loop} & 0 & 0 \\
m_{loop} & 0 & M_N & 0 \\
0 & M_N & 0 & Y_{NS^i}v_{\phi} \\
0 & 0 & Y_{NS^i}v_{\phi}  & 0
\end{matrix}\right)
\label{U1Mnu_1}
\end{equation}
in the $\left(\nu_L,N_{L},N_{R}^c,S_{L}^i\right)$ basis. We choose the basis for $S_{L}^i$ in which the linear combination of the four $S_{L}^i$'s that couple to $N_R$ appears in the mass matrix, and the other three orthonormal combinations do not couple to $N_R$. Before $U(1)_D$ symmetry breaking, {\bf $N_{L(R)}$} is a vectorlike fermion with mass $M_N$, and the neutrinos together with $S_{L}^i$ are massless. After $U(1)_D$ symmetry breaking, we get one heavy Dirac fermion, mostly {\bf $N_{L(R)}$}, and most importantly, the neutrino combines with $S_{L}^i$ to  become a Dirac fermion. The eigenvalues, to the leading order in the $m_{loop}\ll M_N,Y_{NS^i}v_{\phi}$ limit, are approximately given by
\begin{align}
\label{eq:mnu1}
m_{\nu}&\approx \frac{m_{loop}}{\sqrt{1+M_N^2Y_{NS^i}^{-2} v_{\phi}^{-2}}}, m_H \approx \sqrt{M_N^2+Y_{NS^i}^2 v_{\phi}^2}
\end{align}
Neutrinos are Dirac at leading order. Since the lepton number is violated softly by two units by the $\mu_{D} \eta_D^2 \phi$ trilinear term, neutrinos become pseudo-Dirac when higher-loop corrections are considered. In the case where the $m_{loop}$ does not provide enough suppression for the neutrino masses, the ratio of $M_{N}/v_{\phi}$ can provide extra suppression for the neutrino masses. For example, if $m_{loop}\sim10^{-4}$ GeV, then $M_{N}/v_{\phi}\sim 10^{6}$ would give $m_{\nu}\sim O(0.1$ $\mbox{eV})$. The other three $S_{L}^i$ states orthonormal to the $S_{L}^j$ state coupled to the $N_R$ fermion obtain their masses radiatively through $C_L^i$ Majorana masses.

It is worthwhile to note that in the $\mathbb{Z}_2^1\sim\left(-\right)$ odd sector, $A_L$ remains massless till neutrinos get their masses, such that $A_L$ mass is generated through Dirac neutrino mass ($m_{loop}$). To generate radiative neutrino masses, we need a fermion mass in the loop, which is $M_{F_{2}}$ the mass of $F_{2L}$ mass eigenstate. $F_{2L}$ is mixed state of $A_L$ and $C_L$ interaction eigenstates, mostly $C_L$, whereas the other mass eigenstate, $F_{1L}$ mostly $A_L$, is still massless. Radiative neutrino masses are generated without $M_{F_1}$. Therefore, the dominant contribution to the $m_A(M_{F_1})$ comes from the effective four-loop diagram proportional to $m_{loop}$, and $m_{A}$ is then given by
\small{
\begin{align}
\label{eq:MA1}
&\frac{Y_N m_{loop}Y_L}{16\pi^2}\left[s_{\eta R} c_{\eta R} F(x_{1R},x_{2R})+s_{\eta I} c_{\eta I} F(x_{2I},x_{1I})\right],
\end{align}
}
where $F(x_i,x_j)$ and {\bf $x_{i(j)}$} are given in Eq.~(\ref{eq:f1}). This is an important point because if $A_L$ obtained its mass in some other way, the neutrinos would generate their masses through $A_L$'s mass, and the kinetic mixing would contribute in the subleading order and be unnecessary. Also, it is important to mention that  this predicts one dark fermion to be naturally lighter than the $m_{loop}$ since its mass is one-loop suppressed with respect to the $m_{loop}$ and overall four-loop suppressed. To avoid $m_A<m_{\nu}$, we can use the freedom of $M_N$. For instance, setting $M_N/v_{\phi}\sim 10^6$, $m_{loop}\sim10^{-4}$ GeV, and Yukawa's of $O(1)$ would give $m_{\nu}\sim O(0.1$ eV$)$ and $m_A\sim O($ keV$)$. The five copies of $C_L^i$ dark fermions obtain their masses through $\left\langle\phi\right\rangle$ at tree level by incorporating the diagonalization of the $5\times 5$ mass matrix in the $C_{L}^i$ basis. The $\mathbb{Z}_2^2\sim\left(-\right)$ odd sector vectorlike fermion $\Psi$ has an invariant mass of $M_{\Psi}$.

Gauge Bosons: The $U(1)_D$ dark gauge boson gets its mass through a canonical Higgs mechanism during spontaneous symmetry breaking of $U(1)_D$ gauge symmetry in the dark sector. Mass of the dark $U(1)_D$ gauge boson is given by $m_{A_D}^2=2g_D^2v_{\phi}^2$, and the corresponding would-be Nambu-Goldstone boson is Im$[\phi]$. Because of the absence of scalars with nonzero VEV that simultaneously transform under $G_{SM}$ and dark $U(1)_D$ gauge symmetry, there is no tree level mixing between $A_D^{\mu}$ and SM neutral gauge bosons. Mixing will appear at one-loop order through $\eta_L^{\pm,0}$ running in the loop but it is loop suppressed and we will ignore the mixing here. The rest of the gauge bosons obtain their masses just like in the SM.

Scalars: The charged Higgs scalar from the SM $H^{\pm}$ corresponds to the would-be Nambu-Goldstone boson and gets eaten up by $W^{\pm}$. The other electrically charged scalar $\eta^{\pm}_L$, which is part of the $\eta_L$ doublet needed for the neutrino mass generation, does not mix with $H^{\pm}$ due to the presence of $\mathbb{Z}_2^1$ under which $\eta_L\sim -$ and $H\sim +$. For electrically neutral scalar components, we have three sectors: the mixing of real components of $H^0$ and $\phi$, and the real and imaginary components of $(\eta_L^0, \eta_D)$ and $(s_{7}, s_{11})$. $(H^0_I, \phi_I)$  correspond to the longitudinal degrees of freedom of the $Z$ and $A_D$ gauge bosons. The three separate sectors arise due to the 2$\times\mathbb{Z}_2$ symmetries present; the first one is analogous to the canonical scotogenic model and plays the same role here, whereas the second $\mathbb{Z}_2$ symmetry in this case is unique and is present here due to the kinetic mixing mechanism and the fractional charges of the particles involved. It can be thought of as a dark stabilizing symmetry for the dark sector within dark sector of the scotogenic model. In this way, the corresponding neutral scalars can be categorized as $(H^0,\phi)\in\{+,+\}$, $(\eta_L^0,\eta_D)\in\{-,+\}$, and $(s_{7},s_{11})\in\{+,-\}$ under the two $\mathbb{Z}_2$ symmetries. All $2\times 2$ scalar mass matrices can be easily diagonalized using unitary rotations with one angle.

The $G_{SM}\otimes U(1)_D$ model can accommodate the multicomponent DM scenario. The lightest of the  particles that transform as $\mathbb{Z}_2^{1,2}\sim(-,+)$ is one component, and the stability is provided by the $\mathbb{Z}_2^1$ symmetry, which is exact. Assuming $M_{\Psi}>m_{s_{7,11}}$, the second component is the lightest eigenstate of the $s_{7},s_{11}$ sector, which transforms as $\mathbb{Z}_2^{1,2}\sim(+,-)$. In this case, $\Psi\sim(-,-)$ under $\mathbb{Z}_2^{1,2}$ would decay into lighter $s_{7,11}$ eigenstates$\sim(+,-)$ and $A_L\sim(-,+)$ through $Y_{A,C}$ Yukawa couplings.

In this paper, we have presented neutrino mass generation via a fermion kinetic mixing mechanism in the context of the anomaly free $SU(3)_c\otimes SU(2)_L\otimes U(1)_Y\otimes U(1)_D$  gauge symmetry. In this case, neutrino Dirac mass is generated after EW and $U(1)_D$ symmetry breaking via the kinetic mixing of two fermions in the dark sector. As a consequence, the neutrino mass is naturally suppressed by the radiative nature of the generation mechanism. This model includes two dark sectors: The first one is similar to the scotogenic scenario, and the second is unique to realize the kinetic mixing mechanism, which allows for the multiparticle DM scenario.
Despite presenting the particular example with $G_{SM}\otimes U(1)_D$ gauge symmetry, the kinetic mixing idea is more general and can be realized in cases with other gauge symmetries as well. In principle, the kinetic mixing of fermions does not need to be carried out in the dark sector, and we could kinetically mix neutrinos with other fermions, but this would require us to include sterile neutrinos in the model. Furthermore, in this case, in order to increase the neutral fermion mass matrix rank(to give the neutrino mass), one would need to follow the following scenario: The neutrino mixes with another neutral fermion leading to mass generation of the dark sector. Then using the same diagram, the neutrino would get mass through the mediation of the same dark sector. These and other prospects and phenomenology are among the further possible research directions of this work.

\acknowledgments
This work was supported by the National Research Foundation of Korea Grants No. 2009-0083526, No. 2017K1A3A7A09016430, and No. 2017R1A2B4006338.

\bibliography{references}
\end{document}